\documentstyle[emulateapj]{article}

\begin{document}

\submitted{Submitted to ApJL}

\title{Cosmic Shear and Power Spectrum Normalization with the 
Hubble Space Telescope}

\renewcommand{\thefootnote}{\fnsymbol{footnote}}
\author{Alexandre Refregier$^{1}$, Jason Rhodes$^{2}$\footnote[2]
{NASA/National Research Council Research Associate},
\& Edward J. Groth$^{3}$}

\affil{1 Institute of Astronomy, Madingley
Road, Cambridge, CB3 OHA, U.K.; ar@ast.cam.ac.uk}
\affil{2 Laboratory for Astronomy and Solar Physics, Code 681, Goddard Space Flight Center, Greenbelt, MD 20771; jrhodes@band1.gsfc.nasa.gov}
\affil{3 Physics Department, Princeton University,
Jadwin Hall, P.O. Box 708, Princeton, NJ 08544;
groth@pupgg.princeton.edu} 

\begin{abstract}
Weak lensing by large-scale structure provides a direct measurement of
matter fluctuations in the universe. We report a measurement of this
`cosmic shear' based on 271 WFPC2 archival images from the Hubble
Space Telescope Medium Deep Survey (MDS). Our measurement method and
treatment of systematic effects were discussed in an earlier paper.
We measure the shear variance on
scales ranging from 0.7' to 1.4', with a detection significance
greater than 3.8$\sigma$. This allows us to measure the normalization
of the matter power spectrum to be $\sigma_{8} = (0.94 \pm 0.10 \pm
0.14) (0.3/\Omega_m)^{0.44} (0.21/\Gamma)^{0.15}$, in a $\Lambda$CDM
universe. The first $1\sigma$ error includes statistical errors only,
while the latter also includes (gaussian) cosmic variance and the
uncertainty in the galaxy redshift distribution. Our results are
consistent with earlier cosmic shear measurements from the ground and
from space. We compare our cosmic shear results and those from other
groups to the normalization from cluster abundance and galaxy
surveys. We find that the combination of four recent cosmic shear
measurements are somewhat inconsistent with the recent normalization
using these methods, and discuss possible explanations for the
discrepancy.

\end{abstract} \keywords{cosmology: observations --- dark matter
---gravitational lensing --- large-scale structure of the universe}

\section{Introduction}
\label{introduction}
Weak gravitational lensing by large-scale structure has been shown to
be a valuable method of measuring mass fluctuations in the universe
(see Mellier at al. 2001 for a review). This effect has been detected
both from the ground (Wittman et al. 2000; van Waerbeke et al. 2000,
2001; Bacon et al. 2000, 2002; Kaiser et al. 2000; Hoekstra et
al. 2002) and from space (Rhodes, Refregier, \& Groth 2001, RRGII;
H\"{a}mmerle et al. 2001).  These results bode well for the prospect
of measuring cosmological parameters and the mass distribution of the
universe using weak lensing.

In this letter, we present the highest significance detection of cosmic
shear using space-based images. It is based on images from the Hubble
Space Telescope (HST) Medium Deep Survey (MDS; Ratnatunga et
al. 1999). We apply the methods for the correction of systematic
effects and detection of shear we have previously developed (Rhodes,
Refregier, and Groth 2000; RRGI) to 271 WFPC2 fields in the MDS. The
method of RRGI is specifically adapted to HST images and takes
advantage of the small PSF of the HST (0.1'' as compared to
$\sim$0.8'' from the ground).  This affords us a higher surface
density of resolved galaxies as well as a diminished sensitivity to
PSF smearing when compared to ground-based measurements. We develop an
optimal depth-weighted average of selected MDS fields to extract a
weak lensing signal.  We then use this signal to derive constraints on
the amplitude of the mass power spectrum and compare this to
measurements from previous cosmic shear surveys and from other
methods.
 

\section{Data}
\label{data}
The MDS consists of primary and parallel observations taken with the
Wide Field Planetary Camera 2 (WFPC2) on HST. We selected only the
I-band images in chips 2,3, and 4 to study weak lensing. To ensure
random lines-of-sight, we discarded fields which were pointed at
galaxy clusters, leaving us with 468 I-band fields.  To ensure that
our fields are independent, we selected 291 fields separated by at
least 10', beyond which scale the lensing signal drops considerably
(see Figure~2).

We used the MDS object catalogs (Ratnatunga et al. 1999) to determine
the position, magnitude, and area of each object, as well as to
separate galaxies from stars. We used the chip-specific backgrounds
listed in the MDS {\tt skysig} files, which are consistent with
backgrounds calculated using the IRAF task {\tt imarith}.  Not using
object-specific backgrounds necessitated the discarding of another 20
fields with a large sky gradient. Our final catalog thus consisted of
271 WFPC2 fields amounting to an area of about 0.36 deg$^{2}$.

\section{Procedure}
\label{procedure}

The procedure we used for measuring galaxy ellipticities and shear
from the source images is described in detail in RRGI\markcite{rho99}
(1999) (see also RRGII and \markcite{rhot99}Rhodes 1999). It is based
on the method introduced by Kaiser, Squires, and Broadhurst (1995),
but modified and tested for applications to HST images.  The
usefulness of our method was demonstrated by our detection of cosmic
shear in the HST Groth Strip (RRGII).

We correct for camera distortion and convolution by the anisotropic
PSF using gaussian-weighted moments. Camera distortions were
corrected using a map derived from stellar astrometric shifts
(Holtzman, et al., 1995). PSF corrections were determined from HST
observations of four stellar fields These fields were chosen to span
the focus range of the HST as shown by Biretta et al. (2000).
Finally, we derive the ellipticities $\epsilon_{i}$ of the galaxies
and convert them into shear estimates using $\gamma_{i}=G^{-1}
\epsilon_{i}$, where $G$ is the shear susceptibility factor given by
equation~(30) in RRGI.

\vspace{0.2cm}
\centerline{{\vbox{\epsfxsize=3.4truein\epsfbox{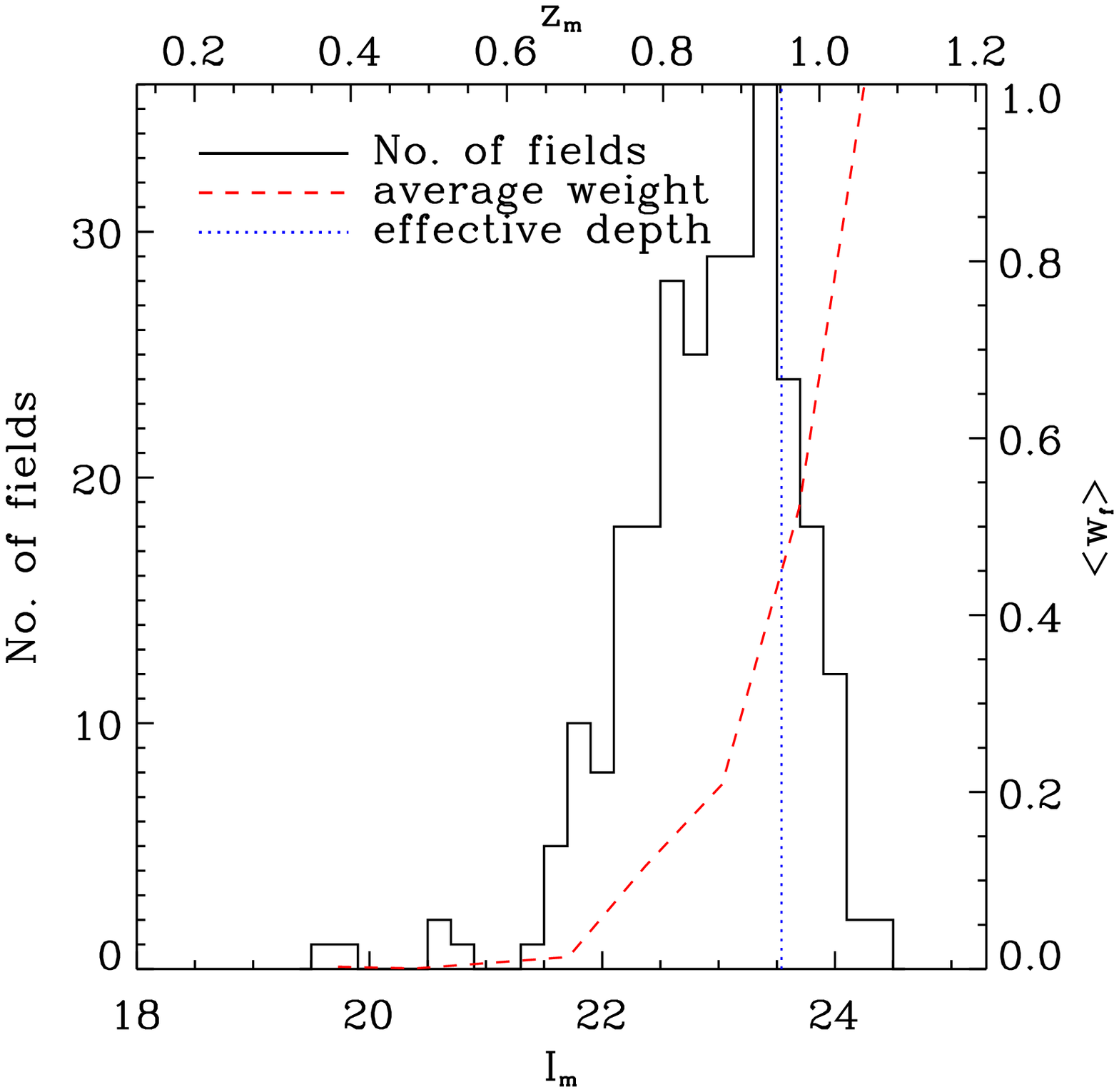}}}}
\vskip-0.3cm \figcaption{\footnotesize Distribution of the median
magnitudes $I_{m}$ of the fields (solid line and left $y$-axis). The
top $x$-axis shows the approximate corresponding median redshift
according to equation~(\ref{eq:red_mag}). The weights of the fields
$\langle w_{f} \rangle$ averaged in each magnitude bin are shown as
the dashed line and right $y$-axis, in arbitrary units. The effective
depth of the survey with this weighted scheme is shown as the vertical
dotted line and corresponds to $I_{m}=23.5$ or $z_{m}=0.95\pm 0.10$.}
\label{fig:magdist}
\vspace{0.2cm}

To limit the impact of noise and systematics, we made a number of cuts
to select our galaxy sample. We first discarded objects for which
$\epsilon>4$ after deconvolution. To avoid using galaxies with low
signal-to-noise, we included only galaxies that have a magnitude
$I<(I_{m}'+2)$, where $I_{m}'$ is the median magnitude (before the
cut) of the field and chip in which the galaxy lies. This is
consistent with the magnitude cut we made in RRGII. We also discarded
small galaxies (size $d<1.5$ pixels) in order to minimize the effects
of the anisotropic PSF on our measurements (see RRGI). The final
galaxy sample contained about $3.1\times 10^{4}$ galaxies. The
distribution of the median magnitude $I_{m}$ (after the cuts) of our
fields is shown in Figure~1, which reveals a wide range of depth for
the MDS.

For the magnitude range of the MDS ($19<I<27$), we use spectroscopic
redshift determinations from the DEEP survey (DEEP Collaboration,
1999) and the Hubble Deep Fields (Lanzetta et al. 1996) to determine
that the median redshift $z_m$ in a field is related to the median
magnitude $I_m$ by the equation
\begin{equation}
\label{eq:red_mag}
z_m \simeq 0.722 + 0.149 \times (I_m - 22.0)
\end{equation}
This agrees well with an extrapolation of the CFRS redshift
distribution (Lilly et al. 1995) we used to determine the median
redshift of objects in the Groth Strip (RRGII). Both methods give
$z_m=0.9 \pm 0.1$ for the Groth Strip ($I_m=23.6$), where the
error gives a measure of the $1\sigma$ systematic uncertainty
in the above relation. 

\section{Estimator for the Shear Variance}
\label{estimator}
We wish to derive a measure of the shear variance on different angular
scales by averaging over the $N_{c}=3$ chips in each of the
$N_{f}=271$ fields. As explained in \S\ref{procedure}, the fields have
varying depths but are are sufficiently far apart to be
statistically independent.  As in RRGII, the total mean shear
$\gamma_{icf}$ in chip $c$ and field $f$ can be measured by averaging
over all the selected galaxies that it contains. It is equal to the
sum of contributions from lensing, from noise and from systematics,
and can thus be written as $\gamma_{icf}=\gamma_{icf}^{\rm
lens}+\gamma_{icf}^{\rm noise}+ \gamma_{icf}^{\rm sys}$. The noise
variance $\sigma_{{\rm noise},cf}^{2} \equiv \langle |\gamma_{cf}^{\rm
noise}|^{2} \rangle$ can be measured from the data by computing the
error in the mean $\gamma_{icf}$ from the distribution of the galaxy
shears inside the chip. As RRGI showed, the systematics are greatly
reduced if the shear is averaged over the chip scale and if small
galaxies (with $d<0.15''$) are discarded. In this case, the
systematics are dominated by the time-variations of the PSF and induce
a shear variance $\sigma_{{\rm sys},cf}^{2} \equiv \langle
|\gamma_{cf}^{\rm sys}|^{2} \rangle$ approximately equal to $0.0011^2$
(see RRGII).

For each field $f$, an estimator for the
shear variance $\sigma_{\rm
lens}^{2}$ on the chip scale is given by
\begin{equation}
\widehat{\sigma^{2}}_{{\rm lens},f} = 
\frac{1}{N_{c}} \sum_{c} |\gamma_{cf}|^{2} -
\sigma_{{\rm noise},f}^{2} - \sigma_{{\rm sys},f}^{2},
\end{equation}
where $\sigma_{{\rm noise},f}^{2} = N_{c}^{-1} \sum_{c} \sigma_{{\rm
noise},cf}^{2}$, and similarly for $\sigma_{{\rm sys},f}^{2}$.
Assuming gaussian statistics, the error variance of the combined
estimator is given by (see Bacon, Refregier \& Ellis 2000 for the case
$N_{c}=2$)
\begin{eqnarray}
\label{eq:sig2_sig2_lens}
\sigma^{2}[ \widehat{\sigma^{2}}_{{\rm lens},f} ] & \simeq &
\frac{1}{N_{c}} \left[ \sigma_{\rm lens,f}^{2} + \sigma_{\rm noise,f}^{2}
+ \sigma_{\rm sys,f}^{2} \right]^{2} \nonumber \\ 
 &  & + \frac{2}{N_{c}^{2}} \sum_{c\neq c'} \sigma_{\times cc'f}^{2}, 
\end{eqnarray}
where $\sigma_{\times cc'f}^{2} = \langle \gamma^{\rm lens}_{1cf}
\gamma^{\rm lens}_{1c'f} \rangle + \langle \gamma^{\rm lens}_{2cf}
\gamma^{\rm lens}_{2c'f} \rangle$ is the cross-correlation between
chips $c$ and $c'$ and can also be measured from the data. The term
$\sigma_{\rm lens,f}^{2}$ corresponds to cosmic variance, and the last
term arises because the chips within a field are not statistically
independent. While the lensing shear field is known to be non-gaussian
on scales smaller than about 10' (eg. Jain \& Seljak 1997), the
non-gaussian corrections to this error estimate are small for
noise-dominated 2-point statistics like the variance (see discussion
in RRGII and White \& Hu 2000).

Because the fields have a range of depths,  it is desirable to combine
the individual estimators $\widehat{\sigma^{2}}_{{\rm lens},f}$ 
using a weighting scheme of the form
\begin{equation}
\label{eq:sigma_lens_wf}
\widehat{\sigma^{2}}_{\rm lens} = \left. \sum_{f}  w_{f}
\widehat{\sigma^{2}}_{{\rm lens},f} \right/ \sum_{f}  w_{f}.
\end{equation}
A convenient choice for the weights is given by
\begin{equation}
w_{f} = \sigma_{\rm noise,f}^{-4},
\end{equation}
i.e. to the inverse square of the noise contribution to the error in
$\widehat{\sigma^{2}}_{{\rm lens},f}$. This weighting scheme is nearly
optimal, and avoids including the lensing signal $\sigma^{2}_{{\rm
lens},f}$ itself. The average weights $\langle w_{f} \rangle$ in
several magnitude bins are shown in Figure~1. As expected, deeper
fields have larger weights since they contain a larger number of
galaxies and thus have a smaller value of $\sigma_{\rm noise,f}$.

To measure the shear variance on the field scale, we first average the
shear within each field and apply the same procedure. This time,
however, the cross-correlation term in
Equation~(\ref{eq:sig2_sig2_lens}) vanishes since each field is
independent. Similarly, we can consider pairs of chips to measure the
shear variance on intermediate scales.

\vspace{0.2cm}
\centerline{{\vbox{\epsfxsize=4truein\epsfbox{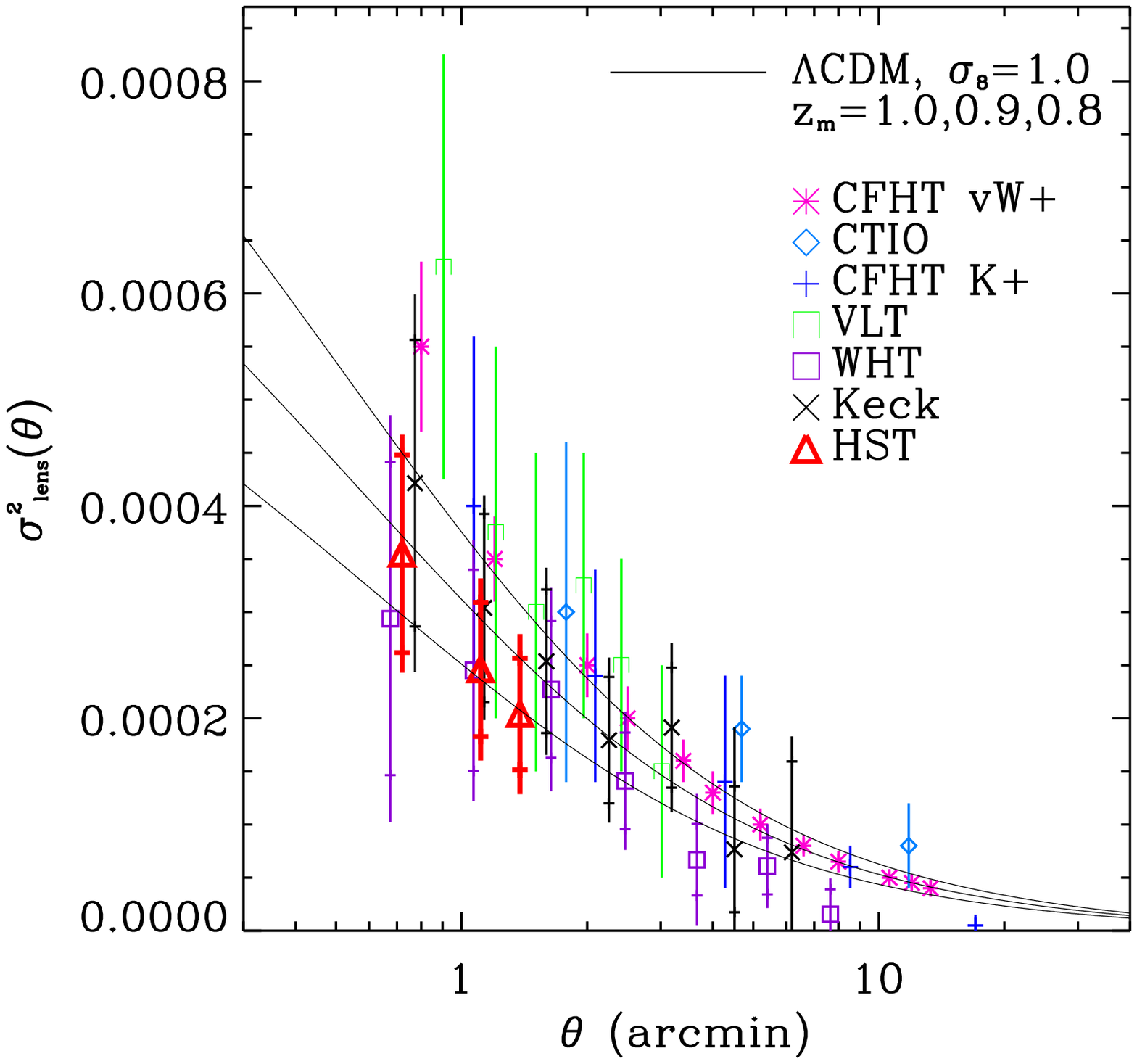}}}}
\vskip-0.3cm \figcaption{\footnotesize Shear variance $\sigma_{\rm
lens}^2$ as a function of the radius $\theta$ of a circular cell. Our
observed value (HST) as well as that observed by other groups: van
Waerbeke et al. (2001, CFHT vW+), Wittman et al. (2000, CTIO), Kaiser,
Wilson \& Luppino (2000, CFHT K+), Maoli et al. (2000, VLT), Bacon et
al.  (2002, WHT and Keck). For our measurement, the
inner error bars correspond to noise only, while the outer error bars
correspond to the total error (noise + cosmic variance). The errors
for the measurements of Maoli et al. (2000) and van Waerbeke et
al. (2001) do not include cosmic variance. The measurements of
H\"{a}mmerle et al. (2001) and Hoekstra et al. (2002) are not displayed
but are consistent with the other measurements. Also displayed are the
predictions for a $\Lambda$CDM model with $\Omega_{m}=0.3$,
$\sigma_{8}=1$, and $\Gamma=0.21$. The galaxy median redshift was
taken to be $z_{m}=1.0,0.9,$ and $0.8$, from top to bottom,
respectively.}
\label{fig:var_obs}
\vspace{0.2cm}

\section{Results}
\label{results}
Our measurement for the shear variance $\sigma_{\rm lens}^2(\theta)$
on different scales is shown in Figure~2. The angular scale $\theta$
is the radius of an effective circular cell whose mean pair separation
equals that of the chip configuration considered ($\theta \simeq
0.72', 1.11'$, and $1.38'$, for 1, 2 and 3 $1.27'$ chips,
respectively). The outer $1\sigma$ error bars include both statistical
errors and cosmic variance (from Eq.~[\ref{eq:sig2_sig2_lens}]), while
the inner error bars only include statistical errors (i.e. by setting
$\sigma_{\rm lens}^2$ and $\sigma_{\times}^{2}$ to 0 on the right-hand
side of this Eq.). For instance, on the chip scale we obtain
$\sigma_{\rm lens}^{2}(0.72')=(3.5\pm 0.9 \pm 1.1)\times 10^{-4}$,
yielding a detection significance (inner error) of $3.8\sigma$ with
this scale alone. As a check of systematics, we analyzed our signal
into $E$ and $B$-modes using the aperture mass $M_{\rm ap}(0.67')$
statistic on the chip scale (Schneider et al. 1998; van Waerbeke et
al. 2001). For $E$-modes, we find the upper limit $\langle M_{{\rm
ap},E}^{2} \rangle =(0.4\pm1.7) \times 10^{-4}$ ($1\sigma$) which is
consistent with the signal expected in a $\Lambda$CDM model ($\langle
M_{{\rm ap},E}^{2} \rangle \simeq 0.6 \times 10^{-4}$; Schneider et
al. 1998).  For the $B$-modes, we find $\langle M_{{\rm ap},B}^{2}
\rangle =(0.3\pm1.7) \times 10^{-4}$ ($1\sigma$), as expected in the
absence of systematics (which corresponds to $\langle M_{{\rm
ap},B}^{2} \rangle \equiv 0$).

The measurements from other groups are also plotted in Figure~2, along
with the predictions for a $\Lambda$CDM model with $\sigma_{8}=1$ and
$\Gamma=0.21$. The central value for $\Gamma$ is close to the recent
measurement of this parameter from galaxy clustering (eg. Percival et
al. 2001), while keeping the $\Gamma=\Omega_{m}h$ relation valid for
$h=0.7$.  The predictions are plotted for a range of galaxy redshifts
$z_{m}=0.9\pm0.1$, corresponding approximately to the uncertainty and
dispersion of this parameter in the different surveys. In our case,
the effective median I-magnitude of our measurement is $I_{m}=\sum_{f}
w_f I_{m,f}/ \sum_{f} w_f \simeq 23.5$, which corresponds to a median
redshift of $z_{m}=0.95\pm0.10$ (see Eq.~[\ref{eq:red_mag}]). The
effective magnitude and corresponding redshift are plotted in
Figure~1.  Given the range of median redshifts in the different
surveys and the correlation between angular bins for the variance, our
results are in good agreement with these other measurements and with
the $\Lambda$CDM model.

Our measurements can be used to constrain cosmological parameters.
Because our measurements on different scales are not independent, we
conservatively only consider the shear variance on the chip scale
($\theta =0.72'$). Within a $\Lambda$CDM model, it is predicted to be
(see RRGII), within an excellent approximation,
\begin{equation}
\label{eq:sigma_lens}
\sigma_{\rm lens} \simeq 0.0202 
 \left( \frac{\sigma_{8}}{1} \right)^{1.27} 
 \left( \frac{\Omega_{m}}{0.3} \right)^{0.56}
 \left( \frac{z_{m}}{0.95} \right)^{0.89}
  \left( \frac{\Gamma}{0.21} \right)^{0.19},
\end{equation}
where $\sigma_{8}$ is the amplitude of mass fluctuations on 8
$h^{-1}$Mpc scales, and $\Omega_{m}$ is the matter density
parameter. Inverting this equation, we find that our measurement of
$\sigma_{\rm lens}^{2}$ yields $\sigma_{8} = (0.94 \pm 0.10 \pm 0.12)
(\Omega_{m}/0.3)^{-0.44} (\Gamma/0.21)^{-0.15} (z_{m}/0.95)^{-0.70}$,
where the first error is statistical only and the second also includes
cosmic variance. To this error must be added that arising from the
uncertainty in the median effective redshift $z_{m}=0.95 \pm 0.10$. After
propagating this error, we obtain
\begin{equation}
\sigma_{8} = (0.94 \pm 0.10 \pm 0.14)
\left(\frac{0.3}{\Omega_{m}}\right)^{0.44} 
\left(\frac{0.21}{\Gamma}\right)^{0.15},
\label{eq:sigma8}
\end{equation}
where the first error reflects statistical errors only, and the latter
is the total error and includes statistical errors, cosmic variance,
and redshift uncertainty (all $1\sigma$).

Figure~3 shows the comparison of our measurement of $\sigma_{8}$
(HST/WFC2) with that from other weak lensing surveys and from other
methods. A $\Lambda$CDM model with $\Omega_{m}=0.3$ and $\Gamma=0.21$
was assumed (except for van Waerbeke et al. 2001 who marginalized over
$\Gamma$). Our $\sigma_{8}$ value is consistent with the other recent
cosmic shear measurements of Bacon et al. (2002), Hoekstra et
al. (2002), and van Waerbeke et al. (2002) and also with the `old'
normalization from cluster abundance (eg. Pierpaoli et al. 2001). This
was recently revised to a lower normalization, by using the observed
mass-temperature relation rather than that derived from numerical
simulations (eg. Seljak 2001). A similar normalization was derived by
combining measurements of galaxy clustering from 2dF and CMB
anisotropy (Lahav et al. 2001). Our results are consistent with this
new normalization at the $1.4\sigma$ level.

\vspace{0.2cm}
\centerline{{\vbox{\epsfxsize=4truein\epsfbox{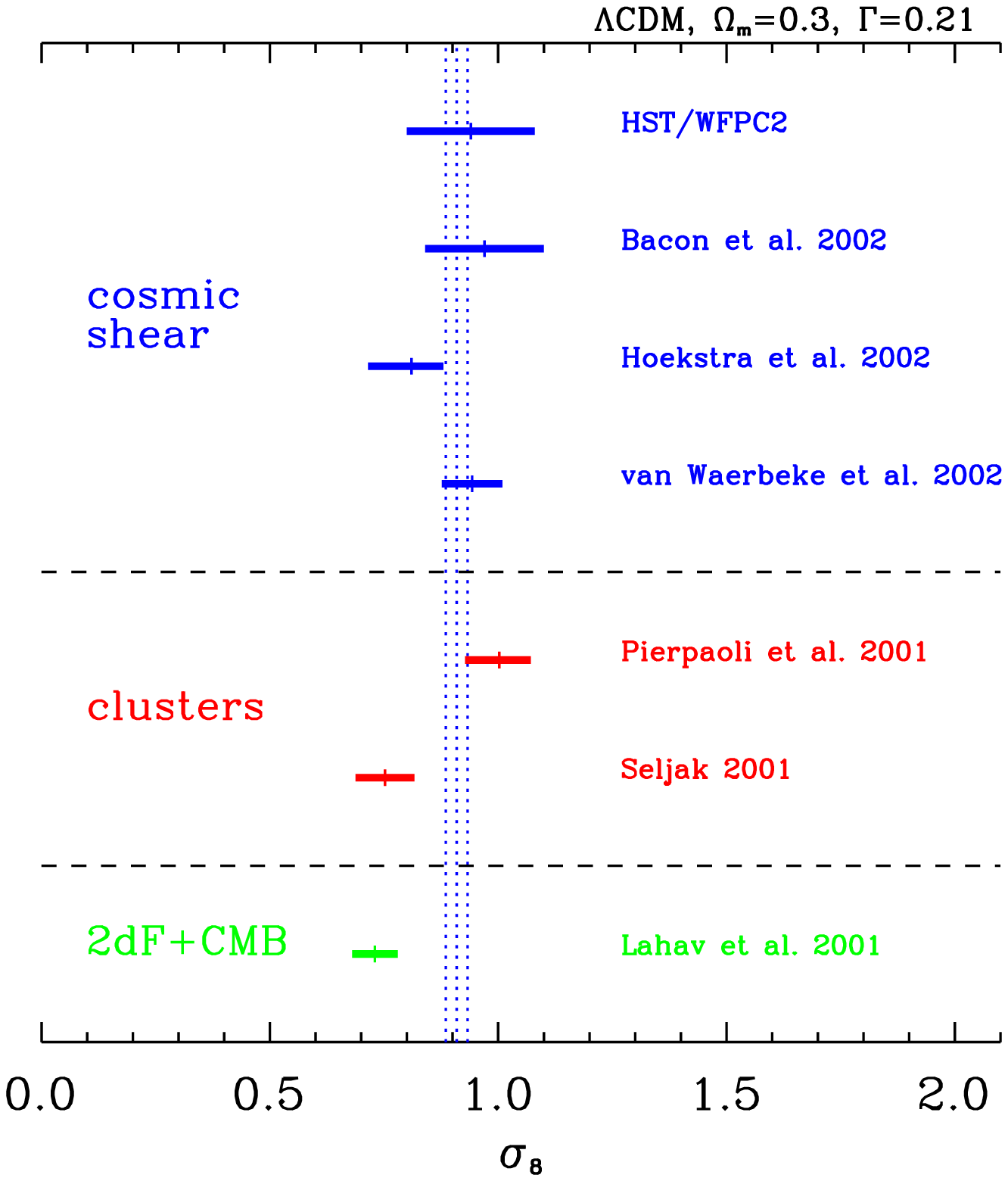}}}}
\vskip-0.3cm \figcaption{\footnotesize Comparison of the determination
of $\sigma_{8}$ by different groups and methods. The errors have all
been converted to $1\sigma$, and a $\Lambda$CDM model with
$\Omega_{m}=0.3$ and $\Gamma=0.21$ was assumed (except for van
Waerbeke et al. 2002 who marginalized over $\Gamma$ between 0.1 and
0.4). The vertical dotted lines shows the weighted average (weights
$\propto \sigma^{-2}$) of the 4 cosmic shear measurements and
associated $1\sigma$ error.}
\label{fig:sigma8}
\vspace{0.2cm}

\section{Discussion and Conclusion}
\label{conclusions}

We have achieved the highest significance detection of cosmic shear
using space-based images to date.  Using the MDS, we have detected the
shear variance on 0.7' to 1.4' scales with a significance greater than
$3.8\sigma$. From the amplitude of the signal we derived a
normalization of the matter power spectrum given by
Equation~(\ref{eq:sigma8}), with errors which include statistical
errors, (gaussian) cosmic variance and the uncertainty in the galaxy
redshift distribution. Our results agree with previous measurements of
the rms shear using both ground and space based images at the
$1\sigma$ level and with the `old' (eg. Pierpaoli et al. 2001) and
`new' (eg. Seljak 2001) cluster abundance normalization at the
$0.4\sigma$ and $1.4\sigma$ level, respectively.

A weighted average of the four recent cosmic shear measurements shown
in Figure~3 yields $\sigma_{8}=0.91\pm0.02$, for a $\Lambda$CDM model
with $\Omega_{m}=0.3$ and $\Gamma=0.21$ (see vertical bars in
Figure~3). This is consistent with the old cluster normalization at
the $1.2\sigma$ level, but somewhat inconsistent with the new cluster
normalization at the $2.5\sigma$ level, where the uncertainty is
dominated by that from cluster abundance. This discrepancy could be
caused by unknown systematics in the cluster abundance or cosmic shear
methods. For the latter case, the calibration of the shear measurement
methods would need to be revisited, in the context of current and
upcoming surveys. The inaccuracy of the calculation of the non-linear
power spectrum and of the halo mass function may also contribute to
the error budget. If confirmed however, this discrepancy could have
important consequences for our understanding of the physics of
clusters, or may require extensions of the standard $\Lambda$CDM
paradigm for structure formation.

\acknowledgments We thank David Bacon, Richard Massey and Richard
Ellis for fruitful discussions, and the anonymous referee for useful
suggestions. AR was supported by an EEC fellowship from the TMR
network on Gravitational Lensing and by a Wolfson College Research
Fellowship.  EJG was supported by NASA Grant NAG5-6279. JR was
supported by an National Research Council-GSFC Research Associateship.
The Medium Deep Survey catalog is based on observations with the
NASA/ESA Hubble Space Telescope, obtained at the Space Telescope
Science Institute, which is operated by the Association of
Universities for Research in Astronomy.

\newpage

\end{document}